# Measuring and Predicting Speed of Social Mobilization

Jeff Alstott[1,2], Stuart Madnick[3], Chander Velu[4]

[1]Section on Critical Brain Dynamics, National Institute of Mental Health, Bethesda, Maryland, USA

[2]Behavioral and Clinical Neuroscience Institute, Departments of Experimental Psychology and Psychiatry, University of Cambridge, Cambridge, UK

[3]Sloan School of Management and School of Engineering, Massachusetts Institute of Technology, Cambridge, MA, USA

[4]Institute for Manufacturing, Department of Engineering, University of Cambridge, Cambridge, UK

Correspondence:    Stuart Madnick, Ph.D., Sloan School of Management, Room E62-422, Massachusetts Institute of Technology, Cambridge, MA, 02142, office: (617) 253-6671, fax: (617) 253-3321; smadnick@mit.edu

**Abstract**

Large-scale mobilization of individuals across social networks is becoming increasingly influential in society. However, little is known about what traits of recruiters and recruits and affect the speed at which one mobilizes the other. Here we identify and measure traits of individuals and their relationships that predict mobilization speed. We ran a global social mobilization contest and recorded personal traits of the participants and those they recruited. We identified how those traits corresponded with the speed of mobilization. Recruits mobilized faster when they first heard about the contest directly from the contest organization, and decreased in speed when hearing from less personal source types (e.g. family vs. media). Mobilization was faster when the recruiter and the recruit heard about the contest through the same source type, and slower when both individuals were in different countries. Females mobilized other females faster than males mobilized other males. Younger recruiters mobilized others faster, and older recruits mobilized slower. These findings suggest relevant factors for engineering social mobilization tasks for increased speed.

**Acknowledgement**:    The authors would like to thank Anton Phillips for operational and financial support as the general manager at Langley Castle, Sunny Cheung for designing and implementing the web site software, and Wei Pan for insights and suggestions based on his experience with the Red Balloon experiment. The authors would also like to thank Roy Welsch and Frank Harrell for their help in statistical interpretation. Contact Stuart Madnick for access to the contest data.

**INTRODUCTION**

Social mobilization is a movement to engage people's participation in achieving a specific goal through self-reliant efforts. Social mobilization can have a broad impact on society, as seen in social movements leading to change in culture or government policy [1–6]. Social mobilization also occurs on much smaller scales, such as friends spontaneously recruiting each other to run search and rescue operations [7,8]. The process of social mobilization can also be purposefully activated and directed; organizations and firms are increasingly turning to social mobilization and large-scale crowdsourcing to solve a variety of problems [9–11]. Using contests as a tool, initial principles have been identified for engineering social mobilization tasks to recruit large numbers of people [12]. Although social mobilization has been studied extensively in a variety of contexts, there has been little attention to measuring the speed of mobilization and quantifying what factors predict that speed. We identify factors that influence the speed of mobilization in recruitment for a contest. These factors may be used to engineer social mobilization tasks so as to recruit people more quickly.

Social mobilization typically spreads on an existing social network. Individuals can activate a single chain of social contacts to connect to a target person, as in the "six degrees of separation" scenario [13–15]. Social activation can also be branching and spread without a target, such as with knowledge propagating through a network [16–20]. Analysis of these phenomena have primarily described the length of the path or the spatial extent of the activity spreading on social networks. However, the details of the speed of propagation can also be critical, particularly in time-sensitive domains like rescue operations or campaigning prior to elections.

For various types of human communication the speed of activity propagation is heterogeneous and its distribution is heavy-tailed [21,22]. Demographic factors influencing speed have also been well-characterized for such passive diffusion-like processes as the spread of product adoption and musical tastes [23–25]. However, in the case of social mobilization, in which individuals are actively recruiting others for a purpose, our understanding of the predictors of speed of mobilization are still at a nascent stage.

Here we show how the personal traits of individuals and the properties of the relationships between individuals in social networks can influence the speed of mobilization. This framework can be considered as a description of the properties of the nodes of the social network (the traits of the individuals; e.g. gender) as well as the properties of the links (the interactions of the individuals' traits; e.g. if both individuals are in the same city). A better understanding of the predictors of social mobilization speed may enable engineering of mobilization scenarios in order to achieve a particular objective. For example, individuals could be organized more quickly towards a task or purpose by targeting the right elements of the social network.

**RESULTS**

We ran a global contest involving time-critical social mobilization, inspired by the Red Balloon Challenge organized by the Defense Advanced Research Projects Agency (DARPA) in the United States in 2009 [12]. The contest was for Langley Castle Hotel in Northumberland, United Kingdom. The task was to find five knights in parks throughout the United Kingdom on a particular weekend, each with an ID code written on their shield. Contest participants registered on the contest website, and could recruit other participants onto their team online in several ways (see Appendix). Participants had financial incentive to form large teams by recruiting new members, who then recruited other members, and so on (example team structure, Fig. 1A). The first registered participant to correctly report the position of a knight was awarded £1,000. The person who recruited that participant received £500, and their recruiter received £250, and so on. This contest incentive structure was previously found to produce large social mobilization [12]. Any team that as a whole found more than one knight would be awarded a £250 bonus, given to the team founder to distribute as desired. Additionally, the team leaders of the first, second, and third largest teams received £1,000, £500, and £250 respectively.

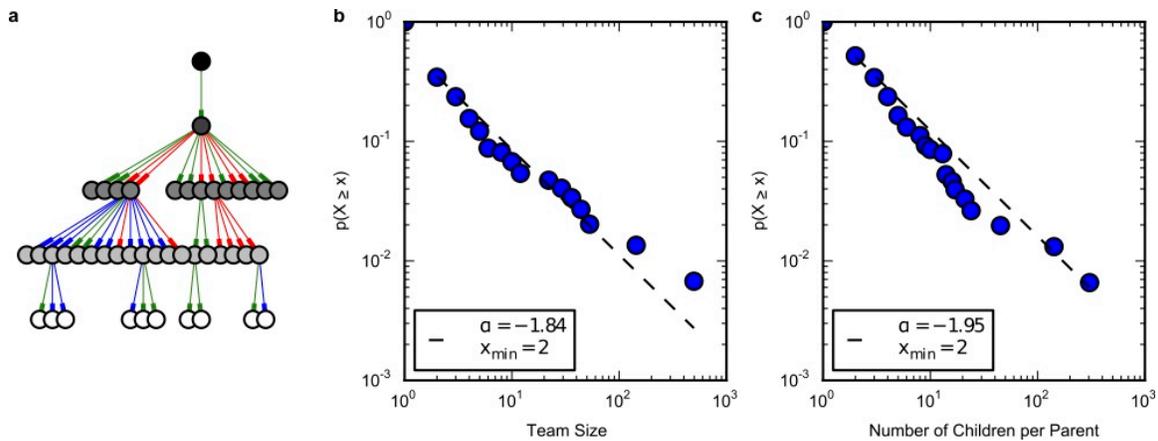

**Figure 1. Mobilized teams grew to a variety of sizes at a variety of rates.**

- **(A)** An example team growing from "parents" to "children", with different parent-child mobilizations having different types of links. The team starter's icon is black, and the future members decrease in shade as their generation in the team increases. Blue links indicate the parent and child heard about the contest through the same type of source (ex. friends). Red links indicate the parent and child heard through different types of sources (ex. family vs. the media). Green links indicate one or both participants did not give information on this trait. This example team was the 4th largest in the contest.
- **(B-C)** Using a similar social mobilization incentive system to that used in the present study, previous research suggested the distributions of team sizes and of parents' number of children followed power laws, with α of 1.96 and 1.69, respectively [12]. We used the statistical methods of Clauset et al. [26] to find only weak to modest support for discrete power laws on these metrics, though with similar values for α. Distribution plots are complementary cumulative distributions (survival functions).
- **(B)** Team size. There were 148 teams, with 51 recruiting additional members beyond the founder. The power law fit was preferred over an exponential (LLR: 58.53, $p<.01$), but was no better of a fit than a lognormal (LLR:.01, $p>.9$)
- **(C)** Number of children for each parent. There were 1,089 participants, with 152 mobilizing at least one child. The power law fit was better than that of an exponential (LLR: 61.45, $p<.02$), but was not a stronger fit than the lognormal distribution (LLR:-.04, $p>.9$)

Unlike the DARPA Red Balloons Challenge that was limited to a single country, two of these knights were "cyber knights", present not in the physical parks themselves but in Google Maps or Google Earth. This allowed for participants outside of the United Kingdom to readily participate, and indeed over 30% of participants with geographic information were from outside the UK.

**Team Creation and Dynamics**

A total of 1,089 participants registered, with 148 starting their own team. Of the teams, 97

did not mobilize any other team members, leaving 51 teams that recruited new participants. In these teams, 152 participants were "parents", mobilizing at least one other participant. These parents mobilized 941 "children". The mean team size was 7.36, and the mean size of teams larger than 1 was 19.45.

To test the robustness of the observed dynamics of this social mobilization contest we compared the size and behavior of the teams to previously reported results from a contest using a similar incentive system [12]. This previous research had suggested the distributions of team size and of parents' number of children both followed power laws. Power laws are very heavy tailed probability distributions, and are notable because they imply the existence of extremely large events, such as a mobilization that grows to encompass the entire global social network. We examined the team dynamics in the present study using rigorous statistical methods [26], described in Methods, and found modest support for power laws, with parameter values consistent with those reported previously (Fig. 1B,C). This replication of previously described team dynamics indicates that at least some features of social mobilization are robust in this style of contest, in which participants recruit others into teams to find particular targets. We now extend the analysis of this type of contest to our primary focus, the speed at which new participants were recruited.

**Measuring and Modeling Mobilization Speed**

When participants registered on the Langley Castle Hotel website to join a team they provided personal information about themselves and how they first heard about the contest, which could be different from the parent who had recruited them onto a team. We also recorded when participants registered on the website. The difference in time between when one participant registered and when a child they recruited registered was the speed of mobilization across that social connection, and is similar to the inter-signup time metric used in Pickard et al. [12]. The mean mobilization speed was 6.7 and the distribution was very heavy-tailed, with a standard deviation of 7.2 days (a histogram of mobilization speeds is shown in Figure 2). There was one month between registration opening and the contest end date, and so the mobilization speed distribution was right-censored; the longest mobilization interval was 26.6 days. The key goal of this study was to uncover the specific factors influencing these mobilization speeds, discussed below.

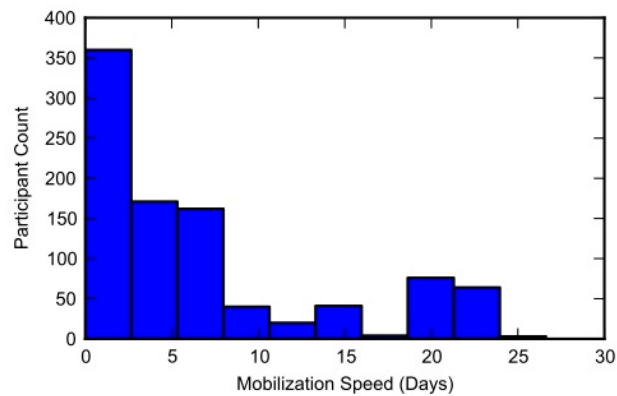

**Figure 2. The distribution of mobilization speeds was heavy-tailed.**

Mobilization speeds were measured by the interval between when a recruiter registered on the contest website and when their recruit registered. The mean mobilization speed was 6.7 days, with a standard deviation of 7.2 days.

We modeled this speed of mobilization with a Cox proportional hazard model, which has been used extensively to describe the spreading of epidemics and subsequently adopted to study diffusion processes on social networks, such as product adoption [23]. The Cox proportional hazard model measured the influence of specific factors on the speed of mobilization, controlling for all other factors. The specific factors examined in our model were: the sources from which participants first heard about the contest, the participants' demographics (age, gender, and location), the number of children the participants recruited, and the timing of individuals' registration relative to both others in their team and the contest timeline.

A hazard function is the likelihood of an event occurring after some time $t$. In our hazard model, the hazard function at time $t$ was the likelihood of a child registering for the contest $t$ units of time after their parent has registered. The influence of a particular factor, such as geographic location, was observed by how much higher or lower the hazard was in the presence of that factor relative to a baseline. This increase or decrease in hazard to baseline was expressed as a hazard ratio. Higher hazard ratios reflected higher likelihoods of registering for the contest at all times $t$, which indicated a faster social mobilization speed. Lower hazard ratios, conversely, indicated slower social mobilization speed, through lower likelihoods of registering for all times $t$.

*Influence of Initial Information Source*: Where the participant first heard about the contest influenced mobilization speed, which could be a source other than their recruiting parent. Mobilization speed increased when the participant heard about the contest direct from the Langley Castle organization (Fig. 3A). From the other categories of information source, the next highest speeds were from family members, then friends, down to the participant's organization or simply the media (difference of hazard ratios between "Langley Castle" and "Media", p<.01; all statistical tests on hazard ratio differences are derived from $\chi^2$ tests). This trend suggests that a direct relationship between the participants and the organizers of the mobilization yields the fastest mobilization speeds. In the absence of a direct exposure to the contest organization, hearing about the contest from more intimate or psychologically close sources of information produced faster social mobilization. Mobilization speed increased when the child and parent first heard about the contest through the same type of source (Fig. 3B).

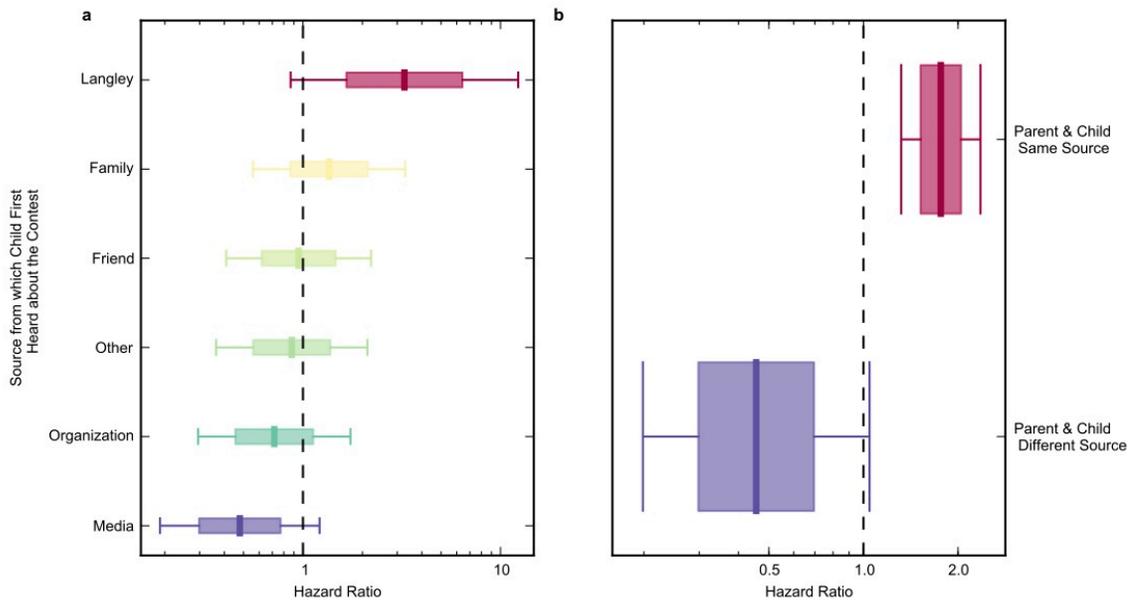

**Figure 3. Where a participant heard about the contest affected mobilization speed.**

In all figures hazard ratios are the increase (>1) or decrease (<1) in likelihood of registering for the contest on a given day, reflecting an increase or decrease in mobilization speed. Boxes represent standard errors, and whiskers represent 95% confidence intervals. Redder boxes indicate faster mobilization (higher hazard ratios), while bluer boxes indicate slower mobilization (lower hazard ratios). Unless otherwise noted, the reference rate (hazard ratio = 1) is for participants who did not give data on that variable, or parent-child pairs in which at least one of the participants did not give data.

**(A)** Mobilization speed was fastest when participants hear about the contest from Langley Castle directly, and decreased as the source is more psychologically distant from the participant.

**(B)** Mobilization was faster when the parent and child heard about the contest through the same category of source than when they heard through different sources.

*Influence of Geography*: Social mobilization speed was also lower when the parent and child were in different countries and fastest when they were in the same city (Fig. 4; p<.01). Our findings show that even in an era of increased telecommunications and "flattening" of the world, geographic proximity still holds considerable influence on how quickly teams mobilize.

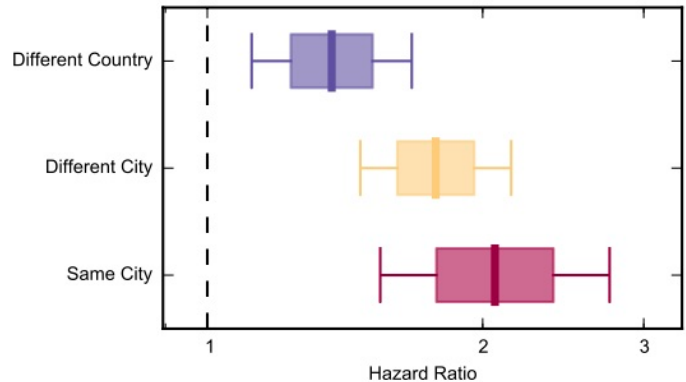

**Figure 4. Geographically closer relationships had faster mobilization speed.**

Social mobilization was faster when the parent and child are in the same city, and slowest when they were in different countries.

*Influence of Gender*: Recent research on the role of gender in the speed of product adoption spread has yielded conflicting findings on whether males or females have greater influence or susceptibility to influence [23,24]. In the present social mobilization task, females mobilized each other faster than males mobilized each other (Fig. 5; $p<.05$). Females mobilizing males and males mobilizing females did not yield statistically significant different speeds ($p>.05$).

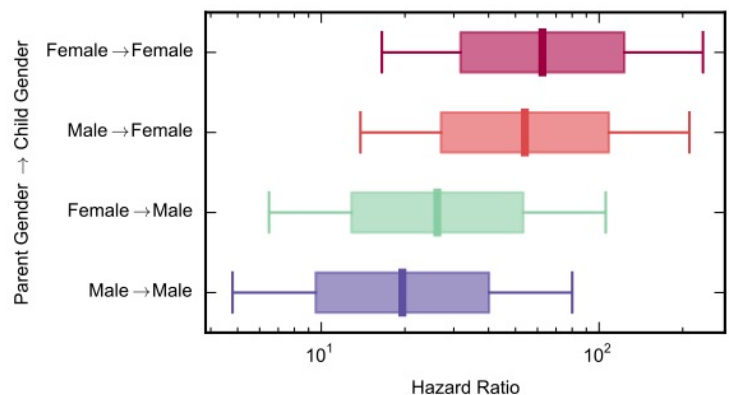

**Figure 5. Females mobilized other females faster than males mobilized other males.**

Females mobilizing males and males mobilizing females did not yield statistically significant different speeds from each other or from the female-female or male-male mobilizations ($p>.05$).

*Influence of Age*: The effect of the parent's and child's ages on mobilization speed were pronounced. Participants' ages were binned into 20-year ranges, and the proportional hazards model included their interaction. For any given parent age group, mobilization speed increased with the child's age. (Fig. 6A). Similarly, for any given child age group, mobilization speed decreased with the parent's age. (Fig. 6C, a rearrangement of the plots in Fig. 6A). These interactions of parent and child age are an instance of the Yule-Simpson paradox [27,28], as they supersede and contrast with the main effects of either child or parent age group alone. Without considering interactions, the main effect of child age showed younger children mobilizing faster (Fig. 6B). Similarly, the main effects of parent age, considered without any interaction with child age, which showed older parents mobilizing faster (Fig. 6D). These main effects mirror findings about age group's influence and susceptibility in passive product adoption spreading on a social network [23]. However, when an interaction effect is present it supersedes that of any main effect, and we show that in social mobilization the role of age is reversed when interaction effects are considered. In particular, young parents and old children display fast mobilization, while old parents and young children display slow mobilization.

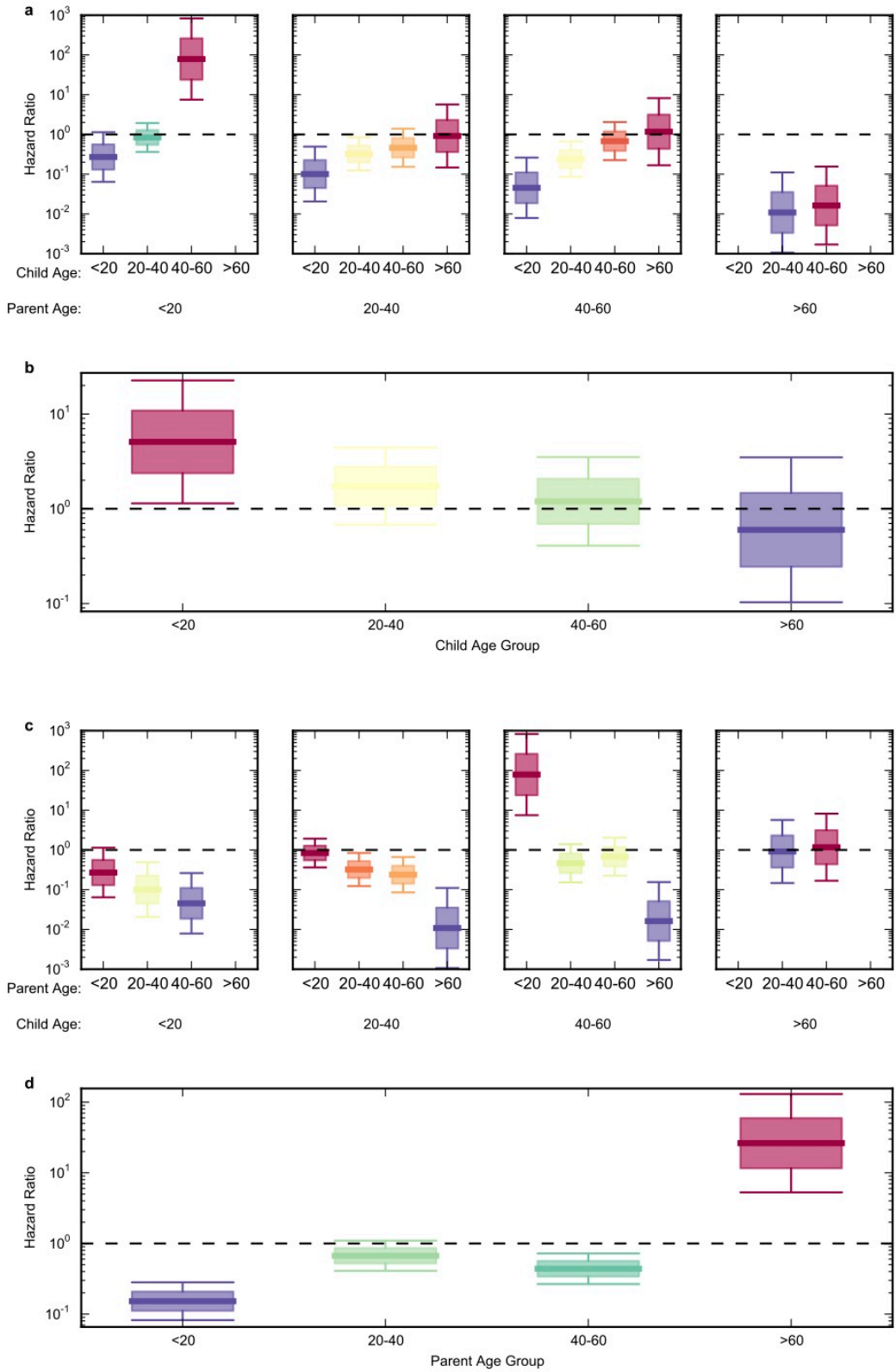

**Figure 6. Older children and younger parents had faster mobilization speeds, as revealed by the interaction of parent and child age.**

In the Yule-Simpson paradox the interaction effect of two factors contrasts with the main effect of either factor taken individually, as is the case with child and parent ages' relationship with mobilization speed. In such a case the interaction effect supersedes the main effect. Absent plots indicate no data for that interaction.

**(A)** The interaction of parent and child age group on mobilization time, grouped by the parent's age. For any given parent age group, mobilization speed increased with the child's age.

**(B)** The main effect of the child's age group on mobilization speed, which had the opposite behavior of that found in the interaction effect seen in (A).

**(C)** The interaction of parent and child age group on mobilization time, grouped by the child age. For any given child age group, mobilization speed decreased with the parent's age. This is a simple rearrangement of the information in (A).

**(D)** The main effect of the parent's age group on mobilization speed, which has the opposite behavior of that found in the interaction effect seen in (B).

*Influence of "Generations":* As teams grew, parents mobilized children, who then in turn became parents mobilizing their own children. This process created "generations" of mobilization within a team. Each additional generation had slower mobilization relative to the one before it (Fig. 7). Lastly, the more future children a participant would have, the faster that participant mobilized. While causality obviously does not allow a participant's number of future children to directly affect his or her own mobilization speed, the statistical relationship indicates that the people that mobilize quickly also recruit more children, independent of other factors.

*Influence of Deadline*: Every additional day after registration opened (meaning one less day until the contest began) the social mobilization speed increased, on average.

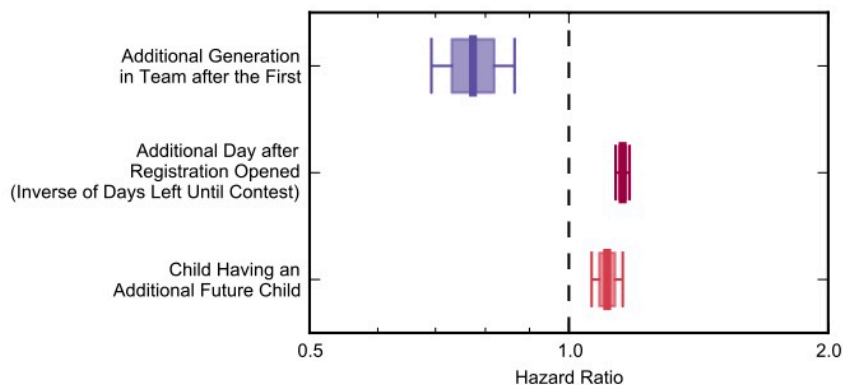

**Figure 7. Additional generations, time left in the contest, and additional future children all affected mobilization speed.**

As a team grew with generations of parents recruiting children, each additional generation beyond the first (hazard ratio = 1) slowed down mobilization speed. In contrast, the further in time the recruiting happened (i.e. closer to the contest date), the faster the mobilization speed. The child's mobilization speed increased for each additional future child he or she had beyond zero.

**DISCUSSION**

As social mobilization becomes increasingly influential, the ability to engineer and influence the dynamics of mobilization will become ever more important within society. Using a contest designed to mobilize a large number of people, we measured the mobilization speed and what factors were associated with the speed of social mobilization. In this contest we found that increased geographic proximity to the mobilizer predicted increased mobilization speed, along with hearing about the contest either directly from the contest organization or from a personal source. Females mobilized other females more quickly than males mobilized other females. Younger people were found to mobilize others with greater speed, and older participants were mobilized more quickly.

These results are from a specific kind of contest, in which individuals self-selected to participate by joining teams for the specific purpose of finding knights for prize money. Controlled experiments in a variety of conditions will be necessary to conclusively show what factors are the determinants of social mobilization speed. However, the present findings provide a preliminary novel approach, analysis, and quantitative understanding that mobilization speed is a function of readily-measurable factors.

Large-scale social mobilizations are becoming increasingly common and impactful, and often the speed of recruitment is critical to their success. A disease prevention campaign, for example, may need to propagate best practices against a new virus quickly. After a natural disaster, donation networks that are set up quickly could provide funds immediately. For those organizing such mobilization tasks, a greater understanding of the factors driving mobilization speed could improve the odds of success. By engineering a few elements of a mobilization task, it could be possible to increase the speed of recruitment. The predictors of social mobilization speed described here compose an initial set of possibly relevant elements, and open the door for identification of additional factors and further research.

**Appendix: Data Collected and Methods of Analysis**

The contest was advertised by Langley Castle, through its web site www.langleycastle.com, newsletters, Facebook pages, email lists, and press releases. A copy of the master press release can be found at [29] and an amusing video is at [30]. Participants registered on the contest website, where they were directed to give demographic information about themselves and how they heard about the contest. Participants could register with their email address and making a password on the site, or alternatively through Facebook Connect. Registration on the website opened June 1, 2011. The competition started on July 2, 2011 at 9am and ended at 9pm on July 3, 2011. The "real" knights were in their assigned parks from roughly 9am to 9pm each day. The "cyber" knights were present on Google Maps and Google Earth all day.

Participants who registered using Facebook Connect could, at the end of the registration process, invite their Facebook friends to join the contest under their team. Registered participants were also provided with a URL they could share with others to register through, which would automatically put those new participants on their parent's team. In addition to the URL, the participants were also given a number that other participants could enter manually to register as their child.

Participants submitted information on knights they found through a form on the website, which required the inclusion of a code unique to each knight printed on their shield. Knights that were already found were announced on the contest website.

**Personal information collected**

The participant's geographic location was inferred from their IP address. Participants whose IP addresses could not be localized to the city level were treated as not having geographic data. The participant registration form's personal information questions and response categories were:

- "Heard about us from": Friend, Family member, Your organization, Langley Castle, Media, Other
- "Gender": Male, Female
- "Age range": < 20 years old, 20-40 years old, 40-60 years old, > 60 years old

Participants were not required to respond on any question. The number of children with available data on a category and the number of parent-child pairings in which both participants had available data on a category were:

- "Heard about us from": 505, 426
- "Gender": 774, 756
- "Age range": 529, 475
- Geographic information: 704, 637

To sageguard personal privacy, participants' data was anonymized before analysis by removal of names, email addresses, and IP addresses. This anonymized dataset is available by contacting the authors.

**Proportional Hazards Model of Mobilization Speeds**

We modeled the speed of mobilization using a single failure, continuous time Cox proportional hazard model. The time modeled was the interval between when a parent and child registered, with the time interval beginning when the parent registered and "failure" when the child registered. The modeled hazard function is the likelihood of a child registering a given time after the parent has registered. This is expressed relative to a baseline hazard function, creating a hazard ratio. Higher hazard ratios reflect higher likelihoods to register at a given time, which when multiplied successively across time speed up the point at which "failure" (registration) will occur. Thus, high hazard ratios indicate fast mobilization speed. Conversely, low hazard ratios extend the time until registration will occur, reflecting slower mobilization speed.

The factors included in the model were:

- the parent's and child's ages and their interaction
- the parent's and child's genders and their interaction
- the type of source from which the child heard about the contest
- if the parent and child heard about the contest from the same type of source
- if the parent and child were in different countries, different cities in the same country, or the same city
- the number of children the parent had
- the number of children the child would have
- the generation the child was in the team
- the time since registration opened (the inverse of the time remaining until the contest began), which was expressed as the date the parent registered

With the child's traits described in a set $X_c$, and the parent's traits described in a set $X_p$, and the child's and parent's ages and genders represented in the set $S(X_c, X_p)$, our model had the form:
$$\lambda_c(t, X_c, X_p) = \lambda_0(t) \exp[\, X_c \beta^c + X_p \beta^p + S(X_c, X_p) \beta^{c,p} ]$$

With $\lambda_0$ as the baseline hazard function as a function of time since the parent's registration $t$, and $\lambda_c(t, X_c, X_p)$ as the hazard for a child at time $t$ with traits $X_c$ and whose parent had traits $X_p$. The coefficient $\beta^c$ is the effect of the child's traits on the hazard, $\beta^p$ is the effect of the parent's traits, and $\beta^{c,p}$ the effect of the interaction of their age and gender traits.

**Power Law Tests of Team Dynamics**

We used the statistical methods of [26] to evaluate whether several features of team dynamics were well-described by power law distributions. These features were the distributions of the number of generations in a team, team size, and a parent's number of children. The statistical methods included using a loglikelihood ratio (LLR) test between a best-fit power law (found through maximum likelihood methods) and an alternative distribution. A positive LLR indicates the power law fit is more likely, and a negative shows the alternative distribution is more likely. The significance of that LLR, however, is given by a p-value. A statistically insignificant LLR means the data does not clearly fit either of the candidate distributions more than the other. Lastly, the best-fit power law may not cover the entire distribution, but only be a good fit beyond a certain value, the $x_{min}$. The shape of these distributions does not impact the use of the Cox proportional hazards model for describing mobilization speed.